\documentclass[aps,prl,twocolumn,superscriptaddress,showpacs,reprint]{revtex4-1}
\usepackage{graphicx, color}				
\usepackage{amsmath, amssymb, amsfonts, mathrsfs}	
\usepackage{times}					
							
\begin{document}

\title{Simulation of Charged Systems in Heterogeneous Dielectric Media via a True Energy Functional}
\author{Vikram Jadhao}
\affiliation{Department of Materials Science and Engineering, Northwestern University, Evanston, Illinois 60208}
\author{Francisco J. Solis}
\affiliation{School of Mathematical and Natural Sciences, Arizona State University, Glendale, Arizona 85306}
\author{Monica Olvera de la Cruz}
\email{m-olvera@northwestern.edu}
\affiliation{Department of Materials Science and Engineering, Northwestern University, Evanston, Illinois 60208}


\begin{abstract}
For charged systems in heterogeneous dielectric media, 
a key obstacle for molecular dynamics (MD) simulations is the need to solve the Poisson equation in the media. 
This obstacle can be bypassed using MD methods that treat the local polarization charge density as a dynamic variable, 
but such approaches require access to a true free energy functional; 
one that evaluates to the equilibrium electrostatic energy at its minimum. 
In this letter, we derive the needed functional.
As an application, we develop a Car-Parrinello MD method for the simulation of free charges present 
near a spherical emulsion droplet separating two immiscible liquids with different dielectric constants.
Our results show the presence of non-monotonic ionic profiles in the dielectric with lower dielectric constant.
\end{abstract}


\maketitle

%
It is hard to overstate the importance of the electrostatic force in biological and soft-matter sciences.
Electrostatic interactions play a major role in determining the structure and function
of several biological macromolecules, such as proteins and DNA \cite{honig,perutz}.
In cell signaling, the creation of electrical potential differences and transfer of ions are of chief importance \cite{clapham}.
On the other hand, electrostatic forces allow the stabilization of many synthetic structures,
endowed with remarkable properties: self-assembled colloidal dispersions \cite{levin1},
overcharged surfaces \cite{charge_inversion}, patterned surfaces by competition between short-range and
Coulombic interactions \cite{paco}, and faceted thin shells \cite{vernizzi}, to name a few.

Investigation of biological and soft-matter systems therefore requires an accurate consideration of electrostatic interactions.
In many situations, computational methods that treat ions individually are necessary.
This is the case, for example, where finite size effects become significant, where the medium has a complex
geometry, or where the dielectric response of the medium is inhomogeneous.
Computing properties of such systems via numerical simulation involves its own challenges.
Due to the long range of the Coulomb force, 
a system of $N$ charges requires an expensive $O(N^{2})$ force calculation at every simulation step.
Attempts to ameliorate this scaling have resulted in the development of several methods:
Ewald summation, particle-mesh methods, fast multipole methods \cite{sagui1},
and the local electrostatics algorithms \cite{sagui2,maggs-rossetto,rottler-maggs}.
In this Letter, we focus on the problems associated with the presence of dielectric heterogeneities in the medium.

The presence of free charges in a medium polarizes its uncharged constituents, leading to a complex
behavior for the electric field and the polarization field itself.
Accurate investigations of systems with electrostatic interactions should incorporate this dielectric response of the medium.
In many cases, it is sufficient to consider the polarization effects by describing the dielectric properties with a spatially varying dielectric constant.
In the simplest case of a uniform dielectric response, a single dielectric constant enters the coarse grained model,
and the simulation proceeds as it would in free space, albeit with a scaled Coulomb's law.
However, most real situations involve regions with different dielectric response, as is the case for
proteins within an aqueous cellular medium or for emulsions where oil and water are partitioned \cite{sacanna}.
In the presence of this varying dielectric response, the simplest form of  Coulomb's law breaks down
and one has to solve the Poisson equation, at \emph{each} simulation step, 
to obtain the necessary force information for the propagation of charges. 
This adversely affects the stability and efficiency of the resulting numerical procedure.
Because of these challenges, the problem of treating variable dielectric response in charge simulations continues
to receive much attention \cite{marchi,allen1,messina1,boda,attard,maggs-rossetto,linse,gan,santos}.

%
In this Letter, we present a variational formulation of electrostatics 
that is applicable to problems involving dielectric heterogeneities.
We construct an energy functional with the polarization charge density as the sole variational field.
This functional works for arbitrary free charge configurations and any kind of spatial variation in dielectric response.
As we review later, these characteristics have not been realized in previously proposed functionals.
We explicitly specialize the functional to the case of sharp interfaces, where only the surface polarization
charge density needs to be considered.
We also demonstrate that our functional can be used in simulations of charged
systems by employing a Car-Parrinello molecular dynamics (CPMD) scheme \cite{car-parrinello} 
where the surface polarization charge density is treated as a dynamical variable.
We note that a similar CPMD approach has been previously proposed \cite{marchi}, 
that employs a functional of the polarization vector in all space as the basic variable. 

We adopt Gaussian units in our formulation. 
The polarization charge density $\omega$ is defined by the relation $\omega = -\nabla\cdot\mathbf{P}$, where $\mathbf{P}$ is the polarization field.
We assume that the medium polarization obeys linear response, $\mathbf{P} = \chi\mathbf{E}$,
where $\chi$ is the susceptibility and $\mathbf{E}$ is the electric field.
Employing the notation $\rho$ for the free charge density and
$G(\mathbf{r},\mathbf{r'}) = |\mathbf{r} - \mathbf{r'}|^{-1}$ for the bare Green's function, our functional reads:
\begin{equation}\begin{split}\label{eq:fnal}
\mathscr{F}[\omega]&=\frac{1}{2}\iint \rho_{\mathbf{r}}G_{\mathbf{r},\mathbf{r'}}
\left(\rho_{\mathbf{r'}}+\Omega_{\mathbf{r'}}[\omega]\right) d^{3}r'd^{3}r\\
&- \frac{1}{2}\iint \Omega_{\mathbf{r}}[\omega] G_{\mathbf{r},\mathbf{r'}}
\left(\omega_{\mathbf{r'}} - \Omega_{\mathbf{r'}}[\omega] \right) d^{3}r'd^{3}r,
\end{split}\end{equation}
where $\Omega_{\mathbf{r}}[\omega]$ is both a functional of $\omega(\mathbf{r})$ and a function of $\mathbf{r}$, and is defined as
\begin{equation}\label{eq:Omega}
\Omega_{\mathbf{r}}[\omega] = \nabla\cdot\left( \chi_{\mathbf{r}}\nabla\int G_{\mathbf{r},\mathbf{r'}}
\left( \rho_{\mathbf{r'}} + \omega_{\mathbf{r'}} \right) d^{3}r'\right).
\end{equation}
Extremization of $\mathscr{F}[\omega]$ leads to the equality: 
$\omega(\mathbf{r}) = \Omega_{\mathbf{r}}[\omega],$
which is the correct physical relation that $\omega$ must satisfy.
In spite of the complex dependence on $\omega(\mathbf{r})$, our functional retains a simple interpretation at equilibrium
as, owing to the extremum condition, its second term vanishes and the first
term becomes the true electrostatic energy
$U=\frac{1}{2}\int\rho(\mathbf{r})\phi(\mathbf{r}) d^{3}r,$
where $\phi(\mathbf{r})$ is the electrostatic potential.
Furthermore, it can be shown that the second variation of $\mathscr{F}[\omega]$ is positive,
implying $\mathscr{F}[\omega]$ becomes a minimum at its extremum. 
We provide the proof of this assertion in the supplemental material. 

A variety of functionals employing various electrostatic quantities
as field variables have been proposed \cite{jackson,marcus,felderhof,radke,karplus,allen1,attard,rottler-maggs,lipparini}.
Many of these functionals are not energy functionals \cite{radke,karplus,allen1,jackson}; namely, they either
become a maximum at extremum \cite{radke,karplus} or evaluate to negative electrostatic energy at equilibrium \cite{allen1,jackson}.
This rules out their use in dynamical optimization methods \cite{car-parrinello}.
Attard \cite{attard} has given an energy functional of the surface polarization charge density, but his functional is derived
for a specific system that involves all free charges to be constrained in one uniform dielectric medium.
Other energy functionals \cite{marcus,felderhof,rottler-maggs} involve relatively expensive vector function variables,
requiring three-dimensional vectorial specification \cite{marchi,rottler-maggs}.
In problems where the dielectric response can be modeled as piecewise uniform, 
our functional reduces to a functional of the surface polarization charge density, which requires only a
two-dimensional scalar specification and offers distinct numerical advantages over the vector variables.

%
Our main result, Eq.~\eqref{eq:fnal}, can be derived as follows.
We begin with the standard expression for the electrostatic energy written as a functional:
$\mathscr{F}\left[\mathbf{E}\right]=\frac{1}{8\pi}\int \epsilon\left(\mathbf{r}\right)\left|\mathbf{E}\left(\mathbf{r}\right)\right|^{2} d^{3}r$,
where $\epsilon$ is the dielectric permittivity.
Next, following \cite{kung},
we include Gauss's law as a constraint to this functional via the Lagrange multiplier $\phi$:
\begin{equation}\begin{split}\label{eq:step1}
\mathscr{F}[\mathbf{E},&\phi]=\mathscr{F}\left[\mathbf{E}\right]
-\int \phi_{\mathbf{r}}\left(\nabla \cdot \left(\frac{\epsilon_{\mathbf{r}}\mathbf{E}_{\mathbf{r}}}{4\pi}\right) -\rho_{\mathbf{r}}\right) d^{3}r.
\end{split}\end{equation}
Note that we take $\mathscr{F}$ to depend parametrically on $\rho$.
Using $\epsilon = 1 + 4\pi\chi$ and $\mathbf{P}=\chi\mathbf{E}$, we introduce $\mathbf{P}$ in \eqref{eq:step1} to obtain
\begin{equation}\begin{split}\label{eq:step2}
\mathscr{F}[\mathbf{E},&\mathbf{P},\phi]=\frac{1}{8\pi}\int |\mathbf{E}_{\mathbf{r}}|^{2} d^{3}r + \int \frac{|\mathbf{P}_{\mathbf{r}}|^{2}}{2\chi_{\mathbf{r}}} d^{3}r  \\
&- \int \phi_{\mathbf{r}}\left( \nabla \cdot \frac{\mathbf{E}_{\mathbf{r}}}{4\pi} + \nabla \cdot \mathbf{P}_{\mathbf{r}} - \rho_{\mathbf{r}}\right) d^{3}r.
\end{split}\end{equation}
Variations of \eqref{eq:step2} with respect to $\mathbf{E}$ and $\phi$ allow us to eliminate all variables in favor of $\mathbf{P}$,
leading to
\begin{equation}\begin{split}\label{eq:step3}
\mathscr{F}\left[\mathbf{P}\right]=\int\frac{|\mathbf{P}_{\mathbf{r}}|^{2}}{2\chi_{\mathbf{r}}}d^{3}r
&+ \frac{1}{2}\iint\left( \rho_{\mathbf{r}} - \nabla\cdot\mathbf{P}_{\mathbf{r}} \right) G_{\mathbf{r},\mathbf{r'}} \\
&\times\left(\rho_{\mathbf{r'}} - \nabla\cdot\mathbf{P}_{\mathbf{r'}} \right) d^{3}r'd^{3}r.
\end{split}\end{equation}

$\mathscr{F}[\mathbf{P}]$ has been derived before using different derivations \cite{marcus,felderhof}.
It can be shown that $\mathscr{F}\left[\mathbf{P}\right]$ is an energy functional; that is, its minimum computes
the equilibrium electrostatic energy \cite{attard}.
We wish to transform $\mathscr{F}\left[\mathbf{P}\right]$ to an energy functional of $\omega$. 
This transition begins by inserting the definition of $\omega$ in
\eqref{eq:step3} via a Lagrange multiplier $\psi$:
\begin{equation}\begin{split}\label{eq:step4}
\mathscr{F}[&\mathbf{P},\omega,\psi]=\int\frac{|\mathbf{P}_{\mathbf{r}}|^{2}}{2\chi_{\mathbf{r}}}d^{3}r
+ \frac{1}{2}\iint\left( \rho_{\mathbf{r}} + \omega_{\mathbf{r}} \right) G_{\mathbf{r},\mathbf{r'}}\\
&\times\left(\rho_{\mathbf{r'}} + \omega_{\mathbf{r'}} \right) d^{3}r'd^{3}r
-\int\psi_{\mathbf{r}}\left( \omega_{\mathbf{r}} + \nabla\cdot\mathbf{P}_{\mathbf{r}} \right) d^{3}r.
\end{split}\end{equation}
We note that $\psi$ can be shown to coincide with the electrostatic potential $\phi$ at equilibrium.
Taking variations of the above functional with respect to $\omega$, $\mathbf{P}$, and $\psi$ gives:
\begin{align}
\begin{split}\label{eq:psi}
\psi_{\mathbf{r}} = \int G_{\mathbf{r},\mathbf{r'}} \left(\rho_{\mathbf{r'}} + \omega_{\mathbf{r'}}\right) d^{3}r',
\end{split}
\\
\begin{split}\label{eq:P}
\mathbf{P}_{\mathbf{r}} = - \chi_{\mathbf{r}}\nabla\int G_{\mathbf{r},\mathbf{r'}}
\left( \rho_{\mathbf{r'}} + \omega_{\mathbf{r'}} \right) d^{3}r',
\end{split}
\\
\begin{split}\label{eq:omega}
\omega_{\mathbf{r}} =
\nabla\cdot\left(\chi_{\mathbf{r}}\nabla\int G_{\mathbf{r},\mathbf{r'}}\left(\rho_{\mathbf{r'}} + \omega_{\mathbf{r'}}\right)d^{3}r'\right).
\end{split}
\end{align}
It is evident from \eqref{eq:psi} and \eqref{eq:P} that $\psi$ is the electrostatic potential.
In addition, following \eqref{eq:Omega},  we can recast Eq.~\eqref{eq:omega} as the equality of two quantities:
the polarization charge density $\omega$ and the operator $\Omega[\omega]$. 
This equality is precisely the extremum condition for $\mathscr{F}[\omega]$.
Substitution of $\psi$, $\mathbf{P}$, and $\omega$ from \eqref{eq:psi}, \eqref{eq:P},
and \eqref{eq:omega} respectively, into \eqref{eq:step4} leads to our central result, the functional in \eqref{eq:fnal}.

We note that not all substitutions lead to our result.
For example, eliminating $\mathbf{P}$ and $\psi$ from \eqref{eq:step4} using \eqref{eq:psi} and \eqref{eq:P} 
gives a functional $I[\omega]$ with the functional density:  
$\rho_{\mathbf{r}}G_{\mathbf{r},\mathbf{r'}}\left(\rho_{\mathbf{r'}}+\Omega_{\mathbf{r'}}[\omega]\right) / 2 
- \omega_{\mathbf{r}} G_{\mathbf{r},\mathbf{r'}}\left(\omega_{\mathbf{r'}} - \Omega_{\mathbf{r'}}[\omega]\right) / 2$.
Upon extremization, $I[\omega]$ singles out the correct physical quantity, but becomes a maximum at equilibrium.
In fact, $I[\omega]$ is exactly the negative of the functional in \cite{allen1}, neither of which are energy functionals.
Functionals $\mathscr{F}[\omega]$ and $I[\omega]$ share a common structure: 
the expression for the total electrostatic energy (the first term in either functionals) 
is constrained by the correct physical relation that $\omega$ must satisfy, 
namely $\omega - \Omega[\omega] = 0$.
The only but crucial difference between these two functionals is the choice of the 
Lagrange multiplier that enforces this constraint.
In past attempts, the Lagrange multiplier that leads to an energy functional 
of $\omega$ remained elusive.
Our current formulation finds the desired multiplier. 

%
We now consider the application of our functional to the problem of point charges in a system with piecewise-uniform dielectrics.
For the sake of brevity, we will restrict ourselves to two uniform dielectrics, 
with different permittivities $\epsilon_{1}$ and $\epsilon_{2}$, separated by a single sharp interface.
Extension to multiple dielectrics and interfaces is straightforward.
We assume that free charges reside only in the bulk of the dielectric.
The free charge density is $\rho(\mathbf{r}) = \sum_{i=1}^{N} q_{i}\delta\left(\mathbf{r}-\mathbf{r}_{i}\right)$, 
where $q_{i}$ is the charge of the $i^{\textrm{th}}$ particle and $N$ is the total number of charges.
For this system, the induced charge density in the bulk has the form
$\omega_{\textrm{bulk}}(\mathbf{r}) = \left(1/\epsilon\left(\mathbf{r}\right) - 1 \right) \rho\left( \mathbf{r} \right)$,
which leads to an effective charge density of $\rho(\mathbf{r})/\epsilon(\mathbf{r})$.
Also, the gradient of $\chi$ vanishes everywhere except at the interface.
Therefore several volume integrals in Eq.~\eqref{eq:fnal} reduce to surface integrals and the 
original functional $\mathscr{F}[\omega]$ is transformed to a functional of the surface induced charge density:
\begin{align}\label{eq:sfnal}
\mathscr{F}[\omega] &= \frac{1}{2}\iint\rho_{\mathbf{r}}K^{^{^{\negthickspace\negthickspace\negmedspace\negthickspace\circ\circ}}}_{\mathbf{r},\mathbf{r'}}\rho_{\mathbf{r'}}d^{3}rd^{3}r'
+ \frac{1}{2}\iint\rho_{\mathbf{r}}K^{^{^{\negthickspace\negthickspace\negmedspace\negthickspace\circ\bullet}}}_{\mathbf{r},\mathbf{s}}\omega_{\mathbf{s}}d^{3}rd^{2}s
\nonumber\\
&+ \frac{1}{2}\iint\omega_{\mathbf{s}}K^{^{^{\negthickspace\negthickspace\negmedspace\negthickspace\bullet\bullet}}}_{\mathbf{s},\mathbf{s'}}\omega_{\mathbf{s'}}d^{2}sd^{2}s',
\end{align}
where $\omega(\mathbf{s})$ is the induced charge density at the position $\mathbf{s}$ on the interface, and
$K^{^{^{\negthickspace\negthickspace\negmedspace\negthickspace\circ\circ}}}$, $K^{^{^{\negthickspace\negthickspace\negmedspace\negthickspace\circ\bullet}}}$, 
and $K^{^{^{\negthickspace\negthickspace\negmedspace\negthickspace\bullet\bullet}}}$ are, respectively, 
the effective potentials of interaction between two free charges, between
a free charge and an induced charge, and between two induced charges. 
These effective interactions are given by:
\begin{align}\label{eq:K}
&K^{^{^{\negthickspace\negthickspace\negmedspace\negthickspace\circ\circ}}}_{\mathbf{r},\mathbf{r'}} = \frac{1}{\epsilon_{\mathbf{r}}}G_{\mathbf{r},\mathbf{r'}}
\,+\, \frac{1}{\epsilon_{\mathbf{r}}}\,\overline{G}_{\mathbf{r},\mathbf{r'}}\,\frac{1}{\epsilon_{\mathbf{r'}}}
\,+\, \frac{1}{\epsilon_{\mathbf{r}}}\,\overline{\overline{G}}_{\mathbf{r},\mathbf{r'}}\,\frac{1}{\epsilon_{\mathbf{r'}}}
\nonumber\\
&K^{^{^{\negthickspace\negthickspace\negmedspace\negthickspace\circ\bullet}}}_{\mathbf{r},\mathbf{s}} = \frac{\epsilon_{\mathbf{r}} - \epsilon_{m}}{\epsilon_{\mathbf{r}}} G_{\mathbf{r},\mathbf{s}}
+ \frac{\overline{G}_{\mathbf{s},\mathbf{r}} - \left(2\epsilon_{m}-1\right)\overline{G}_{\mathbf{r},\mathbf{s}}}{\epsilon_{\mathbf{r}}}
+ \frac{2\overline{\overline{G}}_{\mathbf{r},\mathbf{s}}}{\epsilon_{\mathbf{r}}}
\nonumber\\
&K^{^{^{\negthickspace\negthickspace\negmedspace\negthickspace\bullet\bullet}}}_{\mathbf{s},\mathbf{s'}} = \epsilon_{m}\left(\epsilon_{m}-1\right)G_{\mathbf{s},\mathbf{s'}} -
\left(2\epsilon_{m}-1\right)\overline{G}_{\mathbf{s},\mathbf{s'}} +
\overline{\overline{G}}_{\mathbf{s},\mathbf{s'}}.
\end{align}
Here $\epsilon_{m}=(\epsilon_{1}+\epsilon_{2})/2$ is the permittivity at the interface 
and functions $\overline{G}$ and $\overline{\overline{G}}$ are defined as:
\begin{align}\label{eq:modG}
&\overline{G}_{\mathbf{a},\mathbf{b}} = \epsilon_{d}\int G_{\mathbf{a},\mathbf{u}} \; \hat{n}_{\mathbf{u}}\cdot\nabla_{\mathbf{u}} G_{\mathbf{u},\mathbf{b}} \; d^{2}u
\nonumber\\
&\overline{\overline{G}}_{\mathbf{a},\mathbf{b}}=
\epsilon_{d}^{2}\iint\hat{n}_{\mathbf{u}}\cdot\nabla_{\mathbf{u}} G_{\mathbf{a},\mathbf{u}} \,G_{\mathbf{u},\mathbf{v}}\, \hat{n}_{\mathbf{v}}\cdot\nabla_{\mathbf{v}} G_{\mathbf{v},\mathbf{b}} \,d^{2}ud^{2}v,
\end{align}
where $\epsilon_{d}$ = $|\epsilon_{2}-\epsilon_{1}|/4\pi$ is the permittivity jump at the interface, 
$\mathbf{a}, \mathbf{b}$ are arbitrary position vectors, 
$\mathbf{u}, \mathbf{v}$ are position vectors of arbitrary interfacial points,
and $\hat{n}$ is the unit normal to the interface taken to point in the direction of increasing permittivity.

%
In addition to providing a complete reformulation of electrostatics in heterogeneous media, 
our formalism has immediate applications to important practical problems. 
Since $\mathscr{F}[\omega]$ is an energy functional, 
it can be used for simulating free charges in heterogeneous media which, as described above, 
are basic models for phenomena in both biological and synthetic settings. 
The simplest simulation schemes \cite{allen1,boda} for these systems require, in some way, 
the solution of the extremum condition, $\omega - \Omega[\omega] = 0$, at each step.
However, when an energy functional is available, new approaches are possible, such as the use of CPMD method \cite{car-parrinello}. 
In this approach, $\omega$ is treated as a dynamical variable and is assigned a mass. 
The dynamic equations for the system follow from a Lagrangian that contains an additional kinetic energy term for $\omega$. 
The kinetic term is constructed so that $\omega$ remains close to the exact polarization charge distribution at all times. 
In other words, we replace the expensive solution of the Poisson equation at each simulation step with an {\it on-the-fly} computation of the polarization effects. 

We apply our method to free charges in piecewise-uniform dielectrics, 
where the interfaces between the different uniform dielectrics are closed surfaces of finite area. 
For simplicity, we only consider impenetrable boundaries such that each region has a fixed set of ions. 
We partition the surface boundaries into $M$ finite elements, and to each element $k$ 
we assign an average induced charge density $\omega_k$ and a fictitious mass $\mu_{k}$. 
For the system with $N$ free ions, we can write a Lagrangian for the extended system of $N+M$ particles as:
\begin{equation}\label{eq:lag}
\mathcal{L} =  \sum_{i=1}^{N} \frac{1}{2}m_{i}\dot{\mathbf{r}}^{2}_{i} +\sum_{k=1}^{M}\frac{1}{2}\mu_k \dot{\omega}^{2}_{k} - \mathscr{F}[\omega_{k};\mathbf{r}_{i}]-\mathscr{H}[\mathbf{r}_i].
\end{equation}
The first term is the kinetic energy of $N$ ions with masses $m_i$ and position $\mathbf{r}_i$. 
The second is a fictitious kinetic energy for the surface induced charge density. 
The electrostatic potential energy of the system computed by using our functional constitutes the third term. 
And the final term contains a set of truncated Lennard-Jones potentials 
to model the hard-core of the particles and the impenetrability of the surfaces.

Starting from a point in the extended configuration space, we generate its dynamics via standard MD technique, 
using the equations of motion derived from the Lagrangian $\mathcal{L}$ 
for the ions and the fictitious induced charge values. 
To simulate the behavior of the physical system at finite temperature $T$, 
we couple the augmented system to a set of Nos$\acute{\textrm{e}}$-Hoover thermostats. 
The ions couple to a thermostat at temperature $T$, 
while the induced charge values couple to one at much lower temperature $T_2$. 
This allows the evaluated energy of the physical system to be 
close to its thermal equilibrium value by limiting the contribution of the fictitious kinetic energy. 
This two-temperature approach is a standard feature of CPMD \cite{sprik,blochl-parrinello,fois}. 
The masses of the induced charge degrees of freedom are chosen so as to make their energy contribution small. 
In practice, we choose these masses to be proportional to the areas of the finite elements, and the proportionality constant is chosen to optimize the stability of the simulation. 
Another feature of the system we simulate is that, as a result of Gauss's law, the net induced charge at each interface is a constant. 
In our simulations, we choose to directly enforce this constraint at each step via the SHAKE-RATTLE routine \cite{shake}.

%
\begin{figure}[h]
\centerline{
\includegraphics[scale=0.15]{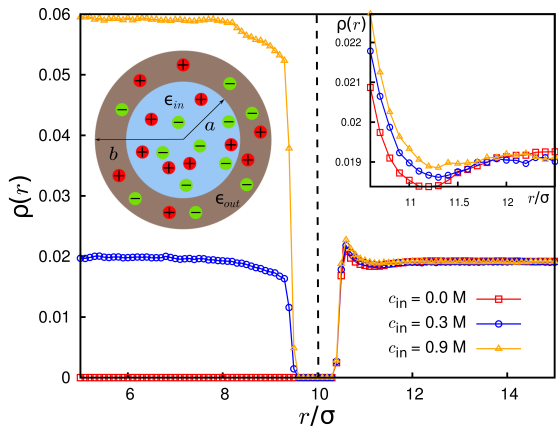}
}
\caption
{\label{fig1}
Ionic density profiles for different concentrations of monovalent electrolyte inside the dielectric sphere. 
The outside salt concentration is held fixed at 0.3 M for all curves.
The inset shows the accumulation of ions near the interface on the lower permittivity side. 
The dashed line in the middle shows the position of the interface.
On the left side we sketch our model for the liquid-liquid emulsion droplet in the presence of a monovalent electrolyte (not to scale).
}
\end{figure}
%
We have used the CPMD method outlined above to simulate ions near a spherical interface separating media of different permittivities. 
The simulations have been carried out for various values of permittivities, ion concentrations and ion valencies.
As a test case, we considered a model for charged colloidal dispersions, where mobile charges are present in only one of the two dielectrics.
We obtained results that match those previously published \cite{messina1,santos}.
In this Letter, we focus on a much less explored problem of ions present in \emph{both} sides of the spherical interface. 
This system can be considered as a model for a liquid-liquid emulsion droplet in the presence of an electrolyte \cite{graaf,bier}. 
We consider the case where the ions do not cross the interface.
We model the ions as repulsive Lennard-Jones (LJ) spheres of diameter $\sigma = 0.357$ nm.
The spherical interface of radius $a=10\sigma$ separates the two media:
the interior dielectric has permittivity $\epsilon_{\textrm{in}} = 80$, while
the exterior dielectric has permittivity $\epsilon_{\textrm{out}} = 35$.
The whole system of ions and dielectric media is taken to be in a spherical simulation cell of diameter $b = 20\sigma$ such that the centers of the 
two spheres coincide, see the sketch in Fig.~\ref{fig1}.
Both the interface and the simulation cell boundary are modeled as repulsive LJ walls.
We consider monovalent electrolyte (1:1) at $T=298$ K and 
$c_{\textrm{in}}$ ($c_{\textrm{out}}$) denote the salt concentrations inside (outside) the spherical interface.
The interface is discretized with nearly $M=2000$ points, and the parameters
associated with the CPMD simulation are: $\Delta=0.001, \mu_{k} \sim 5-10, T_{2} = 0.001T$. Here, $\Delta$ is the simulation 
timestep and these values are given in LJ units (energy: $k_{B}T$, length: $\sigma$).

Our simulation results pass two tests of stability and accuracy. 
We have analyzed the energy of the simulated system as a whole as well as the energies of its subsystems. 
Fluctuations in the total energy of the physical system were found to be 50 times smaller than those in the physical kinetic energy, implying good energy conservation. 
Also, the energy of the fictitious kinetic modes was kept very close to zero at all times. 
Our second test relates to the effectiveness of our scheme to reproduce the correct polarization charge distribution. 
At regular intervals during the course of the simulation of a number of specific cases, we collected and stored the ion coordinates and surface charge densities. 
Then, we carried out an ordinary minimization of the functional to determine the exact induced density, and compared it to the distributions obtained in the simulation. 
Our {\it on-the-fly} method results were within 2\% of those obtained with direct minimization.

Our simulations of the  1:1 electrolyte lead to the density profiles shown in Fig.~\ref{fig1}. 
We consider different values of $c_{\textrm{in}}$ while maintaining $c_{\textrm{out}}$ at 0.3 M, 
thus conducting a study similar in spirit to the experiment in \cite{schlossman1}. 
The ion distributions, for all concentrations, reach a constant value in the bulk on either side but show interesting features near the interface.
On the side having the higher value of permittivity, the ion density is depleted near the interface, 
which is largely the result of the repulsion due to the induced surface charge. 
The depletion is monotonic and gets more pronounced for higher values of $c_{\textrm{in}}$. 
On the other hand, in the exterior, lower-permittivity dielectric, ions prefer to accumulate near the interface (see inset in Fig.~\ref{fig1}).
We also see that ionic profiles on this side of the interface are non-monotonic.
This is due to a combination of Coulombic depletion near hard wall \cite{tsao2,tsao3} and attractive surface polarization charge effects.
We further observe that increasing the internal salt concentration enhances the accumulation of external ions near the interface. 
Several of these features have been previously observed for planar interfaces \cite{boda} and attributed generally to the same basic reasons. 

%
We have presented the solution to the long-standing problem of providing a true free energy functional for 
electrostatics that employs the polarization charge density as the variational field. 
This formulation has many applications, and we have used it to develop an efficient 
CPMD simulation for a set of point charges present in two dielectric media. 
The advantages associated with dynamically optimizing our functional in conjunction with the reduction in 
dimensionality achieved by replacing the polarization vectors with 
induced surface charges paves way for substantial improvements in our ability to simulate charges in heterogeneous media.
As the functional is general for linear media, simulation approaches derived 
from it can be applied to many other systems, such as those with arbitrary interface shapes or moving boundaries.

\begin{acknowledgments}
V.J. thanks R. Sknepnek, P. K. Jha, J. Zwanikken, and G. I. Guerrero-Garc\'{\i}a for valuable discussions.
The authors thank W. Kung for numerous comments on the manuscript.
V.J. was funded by the DDR\&E and the AFOSR under Award No. FA9550-10-1-0167 and F.S. was funded by the NSF Grants No. DMR-0805330 and No. DMR-0907781.
\end{acknowledgments}

\end{document}